\documentclass[prb,aps,twocolumn,showpacs,superscriptaddress]{revtex4}

\usepackage{graphicx}
\usepackage{array}
\usepackage{amsfonts}
\usepackage{amsmath}
\usepackage{amssymb}

\begin{document}
\title{Transport and recombination through weakly coupled localized spin pairs in semiconductors during coherent spin excitation}
\author{V. Rajevac}
\affiliation{Philipps-Universit\"at-Marburg, Fachbereich Physik,
D-35032 Marburg, Germany} \affiliation{Hahn--Meitner--Institut
Berlin, Kekul\'estr. 5, D-12489 Berlin, Germany}
\author{C. Boehme\footnote{electronic mail: boehme@physics.utah.edu}}
\affiliation{University of Utah, Physics Department, 115S 1400E,
Salt Lake City, Utah 84112, USA}
\author{C. Michel}
\affiliation{Philipps-Universit\"at-Marburg, Fachbereich Physik,
D-35032 Marburg, Germany}
\author{A. Gliesche}
\affiliation{Ecole Polytechnique F\'ed\'erale, Institut de
th\'eorie des ph\'enom\`enes physiques, CH-1015 EPF-Lausanne,
Switzerland}
\author{K. Lips}
\affiliation{Hahn--Meitner--Institut Berlin, Kekul\'estr. 5,
D-12489 Berlin, Germany}
\author{S.D. Baranovskii}
\affiliation{Philipps-Universit\"at-Marburg, Fachbereich Physik,
D-35032 Marburg, Germany}
\author{P. Thomas}
\affiliation{Philipps-Universit\"at-Marburg, Fachbereich Physik,
D-35032 Marburg, Germany}
\date{\today}
\begin{abstract}
Semi--analytical predictions for the transients of spin--dependent
transport and recombination rates through localized states in
semiconductors during coherent electron spin excitation are made
for the case of weakly spin--coupled charge carrier ensembles. The
results show that the on--resonant Rabi frequency of electrically
or optically detected spin--oscillation doubles abruptly as the
strength of the resonant microwave field $\gamma B_1$ exceeds the
Larmor frequency separation within the pair of charge carrier
states between which the transport or recombination transition
takes place. For the case of a Larmor frequency separation of the
order of $\gamma B_1$ and arbitrary excitation frequencies, the
charge carrier pairs exhibit four different nutation frequencies.
From the calculations, a simple set of equations for the
prediction of these frequencies is derived.
\end{abstract}


\maketitle

\section{Introduction}
Electrically and optically detected magnetic resonance experiments
(EDMR and ODMR, respectively) are alternative ways to detect
electron spin resonances (ESR) in materials with charge carrier
transport or recombination transitions that are governed by
spin--selection
rules~\cite{Cavenett:1981,stutz,Jiang:2004,Jelezko:2004,KSM,Lep,rong}.
The advantage of EDMR and ODMR in comparison to the traditional
ESR spectroscopy is the sensitivity of these methods which is
typically 6 to 10 orders of magnitude
higher~\cite{Jiang:2004,Jelezko:2004}. This has become
particularly useful for the investigation of paramagnetic centers
in highly diluted matrices or low dimensional semiconductor thin
film devices and interfaces, point
defects~\cite{Cavenett:1981,stutz,boehme:2004,Friedrich:2004,meh}
and defect clusters. One of the challenges of EDMR and ODMR
spectroscopy is that the information obtained from these
experiments is different in comparison to the ESR
data~\cite{lips1}. The reasons for the discrepancies between ESR
and EDMR/ODMR are mainly due to the two different measurement
approaches which imply two different observables: When the density
operator $\hat\rho$ represents the spin ensemble to be
investigated, the observable corresponding to ESR experiments will
always be spin polarization $\langle\vec{P}\rangle=\mathrm{Tr(\hat
P\hat\rho)}$ represented by the spin polarization operator $\hat
P$ whereas for the indirect detection through spin--dependent
transport or recombination, the observables are the
permutation--symmetry or -antisymmetry operators represented by
the singlet $|S\rangle\langle S|$ or triplet operators
$|T_i\rangle\langle T_i|$, respectively~\cite{HabDie}. For many
experimental EDMR/ODMR studies (the so called continuous wave
experiments) the different description of the observables is not
relevant, since these experiments are carried out in the
incoherent time regime where only a line shape analysis of the
respective spectra is feasible. However, when coherent effects are
studied with pulsed techniques
(pEDMR/pODMR)~\cite{Boe7,Boe6,Glasbeek_PhysRevB.19.5549,wrach,schmi}
the interpretation of the experiments strongly relies on the
proper theoretical description of spin interaction during coherent
microwave excitation~\cite{Boe7,boehme:2004}.

An example for the difference between a pEDMR signal and an ESR
signal which come from the same spin ensemble are weakly exchange
and weakly dipolar coupled distant pair states in the band gap of
an arbitrary semiconductor material with weak spin--orbit coupling
as described analytically by Boehme and Lips~\cite{Boe6}. Weak
spin--spin coupling means the coupling is much weaker than the
difference $\Delta\omega=\omega_a-\omega_b$ of the
Larmor--frequencies $\omega_{a,b} =\frac{g_{a,b}\mu_BB_0}{\hbar}$
of the pair partners $a$ and $b$, respectively. Note that $\mu_B$
represents Bohr's magneton and $B_0$ the magnitude of an external
magnetic field to which the spin pair is exposed. Examples for
such systems could be donor--acceptor pairs whose distance is
sufficiently large, yet not large enough to make donor--acceptor
recombination impossible~\cite{spaover}, or equivalently,
trap--dangling bond recombination in disordered silicon materials
such as amorphous or microcrystalline
silicon~\cite{boehme:2004,lips:2004}. Weak spin--orbit coupling is
required in order to ensure spin--conservation and therefore, a
spin--selection rule. It is fulfilled for instance in all known
silicon morphologies but also in many organic semiconductor
materials~\cite{spaover}. When the two pair constituents are
manipulated identically with a coherent pulse of high field
strength~\endnote{Note that throughout this work, the value $B_1$
for the strength of the resonant microwave field is defined as the
strength of the constant magnetic field in the rotating frame
which represents also the amplitude of a circular polarized field
in the laboratory frame. This is in contrast to the definition of
the $B_1$-field in Ref.~\cite{Boe6} which is linearly polarized
along the $\hat x$-axis} $B_1$ ($\frac{g\mu_B B_1}{\hbar}:=\gamma
B_1\gg\Delta\omega$~\endnote{Note that $g$ may be either one of
the two Land\'e--factors $g_a$ or $g_b$ or it may even assumed to
be the free electrons Land\'e factors $g=g_e$ since the
differences between these values in comparison to their magnitude
is negligible here and in all other terms throughout this study
where $g$ appears. For $g=g_e$, $\frac{g\mu_B}{\hbar}:=\gamma$, is
the gyromagnetic ratio.}, whereas $\gamma$ is the gyromagnetic
ratio), they undergo a simultaneous spin--Rabi oscillation. This
can be detected by means of pulse length dependence measurements
with both, pESR and pEDMR. With pESR, the measurement could be
conducted by integration of the free induction decay and would be
called a transient nutation experiment~\cite{schweiger} whereas
with pEDMR, the rate relaxation after the coherent excitation
would be integrated reflecting the pair permutation symmetry
within the pairs at the end of the exciting pulse~\cite{boe:2006}.
While both, the pESR and the pEDMR transients would exhibit
oscillating signals, the frequency of these oscillations would
differ by a factor of 2: The pESR detected nutation frequency
$\Omega_\mathrm{ESR}=\gamma B_1$ would simply represent the Rabi
frequency of an uncoupled spin $s=\frac{1}{2}$, whereas the pEDMR
measured oscillation would exhibit the frequency at which the
identically precessing spins of the two pair partners cross the
geometric plane transverse to the field direction of the
externally applied magnetic field $B_0$ (the $\hat x$-$\hat y$
plane) since at these moments, the projection of the parallel
oriented spins in the $\hat x$-$\hat y$ plane onto the spin
eigenstates with singlet content will be maximized. Since this
plane is passed twice per nutation period, the oscillation of the
transition rate is twice as high. Note that this frequency
discrepancy of the oscillations between pESR and pEDMR detected
transient nutations is changed as the $B_1$--field strength
becomes weak: When $\gamma B_1\ll\Delta\omega$, an ESR excitation
will be possible with only one pair partner at the same time.
Thus, the maximum singlet content of a pair will be achieved when
the spin orientations therein point into opposite directions
parallel to the $B_0$ field. Hence, for weak $B_1$ fields, the
nutation frequencies for pESR and pEDMR become equal.

In the following, a semi--analytical study is presented which
describes spin--dependent electronic transition rates (e.g.
recombination or hopping transport) when the driving forces for
Rabi oscillation (under experimental conditions these are
typically strong coherent microwave fields) are between a weak
($\gamma B_1\ll\Delta\omega$) and a strong excitation regime
($\gamma B_1\gg\Delta\omega$). Goal of this study is to fill the
gap between the two analytically derived extremal cases of very
weak and very strong excitation as presented by Boehme and
Lips~\cite{Boe6} and to describe a general behavior of the
nutation frequency reflected by the spin--dependent transition
rates for arbitrary $B_1$ and $\Delta\omega$ and arbitrary
excitation frequencies $\omega$. For a straight forward
interpretation of experimental results~\cite{Boe:2006b}, it is of
particular interest to understand if the change of Rabi frequency
from $\Omega=\Omega_\mathrm{ESR}$ to $\Omega=2\Omega_\mathrm{ESR}$
takes place continuously or abruptly.
\begin{figure}[b]
\includegraphics[width=84mm]{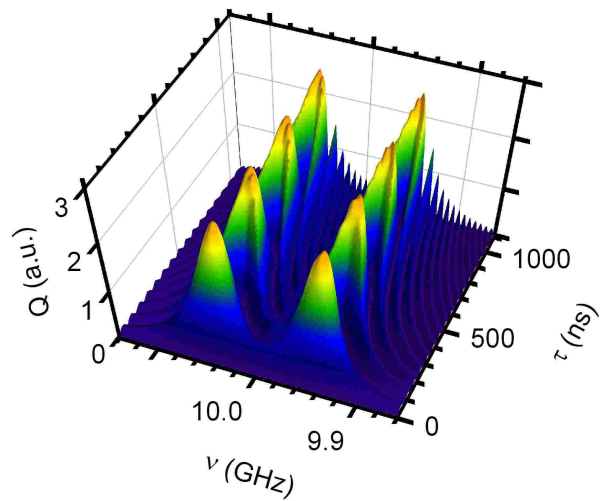}
\caption{Plot of the observable $Q$ as a function of the pulse
length $\tau$ and the applied microwave frequency $\nu$ for a spin
pair with large Larmor separation $\gamma B_1\ll \Delta\omega$.
The parameters of the simulation were
$\frac{\omega_a}{2\pi}=9.95$GHz, $\frac{\omega_b}{2\pi}=10.05$GHz
for $0\leq\tau\leq 1\mathrm{\mu s}$ and microwave excitation
frequencies of $9.86\mathrm{GHz}\leq\nu=\frac{\omega}{2\pi}\leq
10.14\mathrm{GHz}$ as well as a $B_1$-field of $\frac{\gamma
B_1}{2\pi}=10$MHz. One can distinguish the two resonant peaks and
recognize the nutation on the pulse length axis.} \label{fi2}
\end{figure}

\section{Model for spin--dependent recombination}
The basis for the results presented in the following are the pair
models for spin--dependent recombination and transport as
described and discussed in detail in Refs.~\cite{Boe6} and
\cite{boe:2006}, respectively. These models are based on the
Kaplan--Solomon--Mott model~\cite{KSM} under consideration of
non-negligible triplet transition probabilities and spin
relaxation. For the calculation of the data presented we strictly
follow these models under the assumption of negligible spin--spin
interactions. Note that this constraint does not apply to all
known experimental systems and will always have to be considered
when the results presented in the following are applied to the
interpretation of experimental data.

\begin{figure*}
\includegraphics[width=163mm]{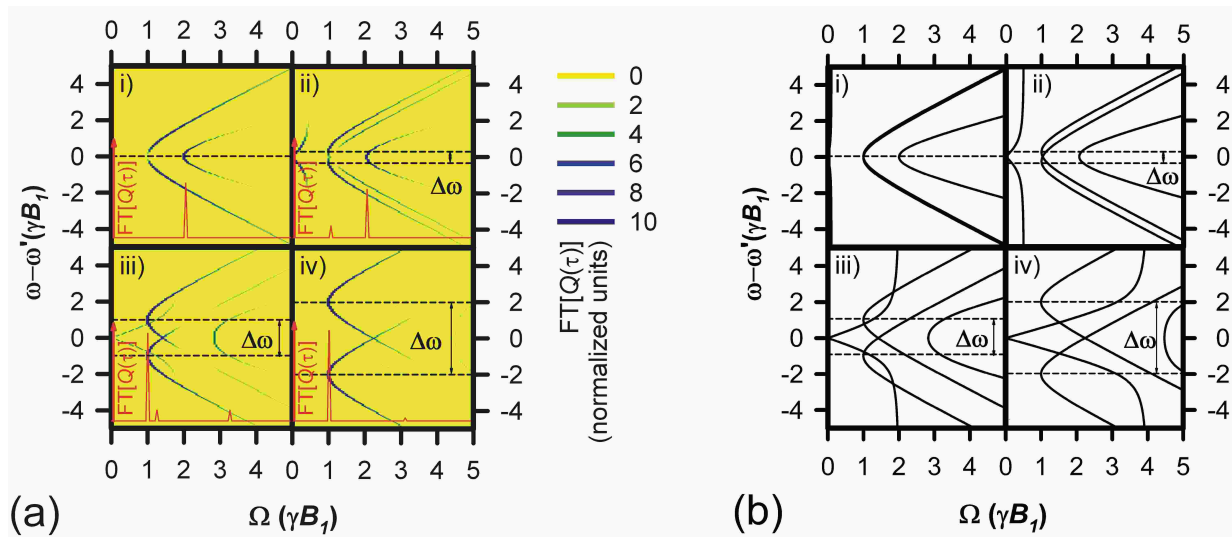}
\caption{(a) Three dimensional color plots of the
semi--analytically calculated $\Omega=\mathrm{FT}\{Q(\tau)\}$ as a
function of the excitation frequency $\omega$ scaled in units of
$\gamma B_1$ as the difference between $\omega$ and the average of
the Larmor frequencies of the two pair partners
$\omega'=\frac{1}{2}(\omega_a+\omega_b)$ and the Rabi--nutation
frequency $\Omega$ in units of $\gamma B_1$. For all four plots
(a.i) to (a.iv), $\frac{\gamma}{2\pi}B_1=10$MHz and
$\frac{\omega'}{2\pi}=10$GHz. From plot (a.i) to (a.iv), the
Larmor separation increases: (a.i),
$\frac{\Delta\omega}{2\pi}=1$MHz; (a.ii),
$\frac{\Delta\omega}{2\pi}=5$MHz; (a.iii),
$\frac{\Delta\omega}{2\pi}=20$MHz; (a.iv),
$\frac{\Delta\omega}{2\pi}=40$MHz. The two dimensional inset plots
display the data of the three dimensional plot at the Larmor
frequency slices indicated by the dashed lines. (b) The plot of
the Rabi--nutation frequencies as obtained from Eqs.~\ref{onespin}
and \ref{twospin} for the same parameters as used for the
calculated results in (a). A comparison with the frequencies
therein shows an excellent agreement.} \label{fi3}
\end{figure*}
Quantitatively, the models outlined can be represented by an
ensemble of spin $s=\frac{1}{2}$ pairs described by the density
operator $\hat\rho$ as derived in Ref.~\cite{Boe6}. We define
$\hat H=\hat H_0+\hat H_1(t)$ to be the Hamiltonian of an
individual pair with $\hat
H_0=-\frac{1}{2}g_a\mu_BB_0\mathbf{\hat\sigma^a_z}
-\frac{1}{2}g_b\mu_BB_0\mathbf{\hat\sigma^b_z}$ representing the
unperturbed Hamiltonian in the presence of a constant magnetic
field $\vec{B}_0=B_0\mathbf{\hat{z}}$ and $ \hat
H_1(t)=-\frac{1}{2} g\mu_BB_1\left(\mathbf{\hat\sigma^a_+}+
\mathbf{\hat\sigma^b_+}\right)e^{-i\omega t} $ is the perturbation
with a circularly polarized microwave of angular frequency
$\omega$ and strength $B_1$. Note the absence of spin--spin
coupling in $\hat H_0$ in contrast to Eq. 5 of Ref.~\cite{Boe6}.
The dynamics of the ensemble of spin pairs can be described by a
Liouville equation $ \partial_t
\hat{\rho}=\frac{i}{\hbar}[\hat{\rho},\hat{H}]^-$ in which, in
contrast to Eq. 1 of Ref.~\cite{Boe6}, all stochastic terms have
been dropped since incoherent processes are considered to be
negligible during the applied coherent ESR pulse. Only coherent
pulses --- these are pulses that are shorter than the fastest
incoherent processes --- are considered in the following. The
ensemble of spins as represented by the density operator
$\hat{\rho}=\hat{\rho}(t)$ can be expressed by a unitary,
time--dependent 4$\times$4 matrix. Due to the absence of coupling,
the Hamiltonian $\hat H_0$ will be diagonal in the product base
with four eigenvalues $\pm\frac{\hbar\omega_0}{2}$ and
$\pm\frac{\hbar\Delta\omega}{2}$ wherein $\omega_0$ and
$\Delta\omega$ represent the sum and the difference of the Larmor
frequencies $\omega_a$ and $\omega_b$, respectively.

When a solution for $\hat\rho(t)$ is found, the transients of the
spin--dependent transition rate
\begin{equation}
R(t)=\sum_{i=1}^4r_i\mathrm{Tr}\left[|i\rangle\langle i|\hat\rho
(t)\right]
\end{equation}
can be calculated from the projection of the permutation symmetry
operators on the ensemble state. Experimentally, a real--time
observation of $R(t)$ on typical time scales in the lower ns-range
is difficult to obtain with pODMR and often impossible with pEDMR
due to the insufficient time resolution caused by the long
dielectric relaxation times of semiconductors in particular at low
temperatures. Hence, transient nutation experiments are typically
conducted by means of decay transient measurements as a function
of the applied pulse length~\cite{Glasbeek_PhysRevB.19.5549}. The
spin--dynamics during the coherent spin excitations are obtained
from these pulse length dependence measurements by charge
integration $Q(\tau)$ which reveals the permutation symmetry state
at the end of the microwave pulse as explained in detail in
Ref.~\cite{boe:2006}. Under the given conditions,
\begin{equation}
Q(\tau)\propto
\Delta(\tau)=\frac{\rho_{11}(\tau)-\rho^S_{11}}{\mathrm{Tr}[\rho^S]}
=\frac{\rho_{44}(\tau)-\rho^S_{44}}{\mathrm{Tr}[\rho^S]}
\end{equation}
whereas $\rho^S_{ii}$ is the steady state value of the matrix
element $\rho_{ii}$ of the density matrix $\hat\rho$. Hence, it is
$Q(\tau)\propto\Delta(\tau)$ which is the observable calculated
and displayed in the following. Note that while $Q(\tau)$
represents a number of charge carriers for pEDMR experiments, the
integration of the photoluminescence decay transient in pODMR
reveals a number of photons. Nevertheless, in both cases, the
observable shall be referred to as $Q(\tau)$ in the following and
is always plotted in arbitrary units because of this ambiguity.

\section{Simulation methods and results}
When incoherence is negligible, the Hamiltonian $\hat H$ can be
diagonalized for any given parameter set and thus, the Liouville
equation in its time integrated form
$\hat{\rho}(\tau)=\exp(-\frac{i}{\hbar}\hat{H}\tau)\hat{\rho^S}
\exp(\frac{i}{\hbar}\hat{H}\tau)$ can be solved by a simple matrix
multiplication in which
$\rho^S=\frac{1}{2}\left[|T_+\rangle\langle
T_+|+|T_-\rangle\langle T_-|\right]$ is the same initial state as
used in Ref.~\cite{Boe6}. The solution $\hat{\rho}(\tau)$ is
referred to as "semi--analytic" since this is easy to perform by
means of calculation of the eigenvalues and eigenvectors of $\hat
H$ with any given parameter set while attempting to solve it fully
analytically for arbitrary variables leads to unreasonably lengthy
expressions.

Figure~\ref{fi2} displays an example for the time domain result
for $Q(\tau)$ as a function of the Larmor frequency for a spin
pair ensemble with a Larmor separation much larger than $\gamma
B_1$ which is one of the extremal cases discussed in
Ref.~\cite{Boe6} (V.A.2.). One can recognize the two well
distinguishable peaks at $\nu=9.95$GHz and $\nu=10.05$GHz on the
Larmor frequency--scale determined by the choice of the respective
$g$-values of the spin--pair partners as well as the undamped
Rabi--nutation on the scale of the pulse length $\tau$. Outside of
the resonances, the signal intensity drops while the nutation
frequency increases. The latter is well known and understood from
the expression of the Rabi frequency as introduced in Eq. 28 of
Ref.~\cite{Boe6}. It is due to the spin nutation about the
residual $B_0$-field $\Delta B_0$ in the rotating frame which
increases with the distance of the excitation frequency from the
resonance frequency. Figure~\ref{fi2} shows that the relevant
information contained in the calculated transients $Q(\tau)$ are
the frequencies as well as the amplitudes of the nutation
components. Thus, from the solutions of $Q(\tau)$ the absolute
Fourier transform (FT) was calculated in order to analyze the
various nutation frequencies contained therein. In contrast to the
short 1$\mu$s transient shown in Fig.~\ref{fi2}, the time scale
for the time domain simulation and, therefore, the Fourier
integration was chosen to be 5$\mu$s long. For the parameters used
throughout this study this corresponds to at least 50 Rabi
oscillation periods.

Figure~\ref{fi3}(a) displays $\Omega$ = FT$\{Q(\tau)\}$ obtained
by semi--analytical calculations for spin pairs with four
different Larmor separations ($\frac{\Delta\omega}{2\pi}$ = 1MHz,
5Mhz, 20MHz, 40MHz) for microwave excitation frequencies of
$9.95\mathrm{GHZ}\leq\nu=\frac{\omega}{2\pi}\leq
10.05\mathrm{GHZ}$ with a given $B_1$-field of $\frac{\gamma
B_1}{2\pi}=10$MHz. While the plots (a.i) and (a.iv) fulfill the
extremal cases of small and large Larmor separation, respectively,
the plots (a.ii) and (a.iii) describe two intermediate cases with
$\gamma B_1\approx\Delta\omega$. Note that the scaling of the
color code was normalized to the maximum for each graph in order
to achieve sufficient contrast. In order to be able to compare the
four different cases more easily, the Rabi components
FT$\{Q(\tau)\}$ at the resonance frequency of the two spin
partners are plotted in the respective graphs (red curves) as
conventional two-dimensional plots. Here, the chosen scaling is
equal for all graphs. The four cases displayed in
Fig.~\ref{fi3}(a) confirm the hyperbolic increase of the Rabi
frequency $\Omega=\sqrt{(\gamma B_1)^2+(\omega-\omega_{a,b})^2}$
as the microwave frequency is shifted out of resonance. One can
deduce from the two--dimensional inset plots of
Fig.~\ref{fi3}(a.i) and (a.iv), that the on--resonance cases show
only one frequency component for the two extremal cases, namely
$\Omega=\gamma B_1$ for large Larmor separation and
$\Omega=2\gamma B_1$ for small Larmor separation. This confirms
the analytical results of Ref.~\cite{Boe6}. Outside of the
resonances ($\omega\neq\omega_{a,b}$), the oscillation splits into
two components for the extremal cases and in the general,
intermediate cases, there are up to four different nutation
frequencies.

\begin{figure}[t]
\includegraphics[width=84mm]{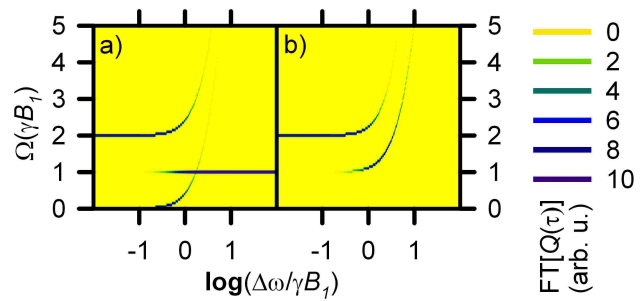}
\caption{Color plot of the Rabi--frequency components
$\Omega=\mathrm{FT}\{Q(\tau)\}$ as function of
$\log\left(\frac{\Delta\omega}{\gamma B_1}\right)$ for an
excitation frequency (a) $\omega=\omega_{a,b}$ on resonance with
one of the pair partners and (b) $\omega=\omega'$ on resonance
with the average $\omega'$ of the pair partners Larmor
frequencies.} \label{fi4}
\end{figure}

In order to illustrate the transition from a single $\Omega=\gamma
B_1$ to a single $\Omega=2\gamma B_1$ frequency component with
decreasing Larmor separation (or equivalently spoken, for an
increasing microwave field $B_1$), a plot of $\Omega$ =
FT$\{Q(\tau)\}$ versus the ratio $\frac{\Delta\omega}{\gamma B_1}$
on a logarithmic scale around $\frac{\Delta\omega}{\gamma B_1}=1$
is shown in Fig.~\ref{fi4} for the one spin on resonant cases
$\omega=\omega_{a,b}$ (a) and the average frequency case
$\omega=\omega'=\frac{1}{2}(\omega_a+\omega_b)$ (b). Again, for
the extremal cases of $\frac{\Delta\omega}{\gamma B_1}\ll 1$ and
$\frac{\Delta\omega}{\gamma B_1}\gg 1$, the plots confirm the
known results for small and large Larmor separation described
above. In Fig.~\ref{fi4}(a), one can see for
$\frac{\Delta\omega}{\gamma B_1}\ll 1$ ($\log
\left[\frac{\Delta\omega}{\gamma B_1}\right] < 0$) that there is
only one component with $\Omega=2\gamma B_1$. As
$\frac{\Delta\omega}{\gamma B_1}$ increases and approaches 1
($\log \left[\frac{\Delta\omega}{\gamma B_1}\right]$ approaches
0), this component gradually becomes weaker and its frequency
increases. This behavior can be understood by the fact that in
Fig.~\ref{fi4}(a), $\omega$ is always equal to one of the pair
partner resonances. When $\Delta\omega$ becomes larger,
$\omega\neq\omega'$ and thus, an increase of the observed nutation
frequency takes place. When $\frac{\Delta\omega}{\gamma
B_1}\approx 1$ ($\log \left[\frac{\Delta\omega}{\gamma
B_1}\right]\approx 0$), two new nutation components become
visible: (i) A low frequency component which generally is hard to
separate from any given offset in the function $Q(\tau)$ and (ii)
one component with $\Omega=\gamma B_1$. The magnitude of the
latter rises from very small values for
$\frac{\Delta\omega}{\gamma B_1}\ll 1$ and increases
asymptotically to a maximum value for $\frac{\Delta\omega}{\gamma
B_1}\gg 1$ whereas the two other components vanish. A
complementary view on these changes is given by Fig.~\ref{fi4}(b)
where the Rabi--components for an excitation frequency
$\omega=\omega'$ are plotted versus $\frac{\Delta\omega}{\gamma
B_1}$. For $\frac{\Delta\omega}{\gamma B_1}\ll 1$, plot (a) and
(b) agree since they represent the same physical situation. When
$\frac{\Delta\omega}{\gamma B_1}$ increases and
$\omega_{a,b}\neq\omega'=\omega$, no low frequency component
becomes visible. The $\Omega=\gamma B_1$ component, which also
becomes visible will increase proportionally to $\Delta\omega$
since it is off-resonant to the applied microwave frequency
$\omega$.

\section{Discussion}
The simulation of spin--Rabi oscillation as observed by
pEDMR/pODMR reveals that the doubling of the nutation frequency
$\Omega=\gamma B_1$ to $\Omega=2\gamma B_1$ with decreasing Larmor
separation ($\frac{\Delta\omega}{\gamma B_1}$) is abrupt which
means that there is no continuous increase of the oscillation
frequency. Instead, only the magnitudes of the various components
change in the intermediate Larmor separation regime about
$\frac{\Delta\omega}{\gamma B_1}\approx 1$. Here, four nutation
components become visible which can all become significant at the
same time. Mathematically, it is obvious that the pairs consisting
of two $s=\frac{1}{2}$ particles with four eigenstates will reveal
four eigenfrequencies for any given general set of parameters. We
interpret this behavior for the general case as the interplay of
the two one--spin systems and the one two--spin system by which
the spin pairs are described in the two extremal cases discussed
above. The results of the analytical derivation given in
Ref.~\cite{Boe6} showed that the most significant qualitative
change that takes place when the Larmor separation changes from
infinity to zero, is that a transition from a one--spin to a
two--spin system occurs. For the one--spin system, the spin in
resonance (note that here, only one spin can be in resonance)
fully determines the oscillation of the pairs permutation symmetry
whereas for the two--spin system, the permutation symmetry is
determined by the relative movements and spin orientations within
the pair ensemble. For large Larmor separation, at least one spin
will remain fixed in its initial state, typically an eigenstate
with polarization parallel to the $B_0$-axis. As the excitation
frequency is changed, the system becomes off-resonant, and two
one--spin contributions
\begin{equation}
\Omega_{a,b}=\sqrt{(\gamma B_1)^2+(\omega-\omega_{a,b})^2}
\label{onespin}
\end{equation}
appear where the frequency of one increases whereas the frequency
of the other decreases at the same time. The pair still behaves
like two individual one--spin systems. For small Larmor
separation, there are always two spins in motion as long as the
system is on resonance (there is then only one resonance line
observed). Hence, since the relative spin motion of the two spins
within the pair will now determine the oscillation of the
electronic rate transition, the beat frequencies
\begin{equation}
\Omega_{p,n}=\Omega_{a}\pm\Omega_{b} \label{twospin}
\end{equation}
of the two one--spin nutation frequencies can be expected. Here,
$\Omega_p$ and $\Omega_n$ stand for parallel and antiparallel
orientations, respectively. Figure~\ref{fi3}(b) displays four
plots of the nutation frequencies obtained with the simple terms
given in Eqs.~\ref{onespin} and \ref{twospin} for the same
parameter sets used for the simulation results displayed in
Fig.~\ref{fi3}(a). While this purely phenomenological description
of the nutation frequencies can not account for the intensities of
the nutation components --- this is the reason why there is no
grey scale gradient in Fig.~\ref{fi3}(b) --- it nevertheless
shows, that there is an excellent agreement of Eqs.~\ref{onespin}
and \ref{twospin} with the semi--analytically calculated frequency
patterns displayed in Fig.~\ref{fi3}(b).

While simple quantitative predictions can be made for the nutation
frequencies, there are no straight forward formulas for the
prediction of the nutation amplitudes. As shown in
Fig.~\ref{fi3}(a), the intensities of the different nutation
components can have quite complex microwave frequency ($\omega$)
dependencies. Well separated lines (case of large Larmor
separation) exhibit Lorentzian line shapes determined by the $B_1$
field due to the power broadening as described in
Ref.~\cite{Boe6}. For intermediate cases, the $\omega$--dependence
becomes much more complex and a given nutation component can
exhibit a maximum at its corresponding resonance but also a local
minimum at the resonance of a non corresponding transition. An
example for this behavior is the data of Fig.~\ref{fi3}(a.iii). At
a frequency of $\omega-\omega'=\pm\gamma B_1$ one can recognize
both, a maximum of the nutation component at $\Omega=\gamma B_1$
but also a minimum of the respective other nutation component at
$\Omega\approx 2\gamma B_1$. Qualitatively, this behavior can be
interpreted by consideration of a four--level system. Any of the
four levels can undergo first order transitions into two different
states. For excitations which are out of resonance with both
transitions, the transition probability is small but may not be
negligible in the vicinity of the resonances. However, when one
transition is induced resonantly, the transition into the
non--resonantly excited state is quenched at the same time and the
intensity of its corresponding nutation component is quenched.
Note that in spite of this qualitative interpretation,
quantitative predictions have to be made by means of the
simulation methods described above.

\section{Summary and Conclusions}
In summary, the response of charge carrier transport and
recombination rates through localized electronic states in
semiconductors to a coherent manipulation by magnetic resonance
were simulated as they would be expected in pEDMR/pODMR
experiments. The transient response was calculated with the spin
excitation present for different excitation fields and frequencies
as well as different Larmor separations within the pairs. It was
assumed that exchange and dipolar interaction are weak and
incoherence due to the electronic transitions or spin-relaxation
is negligible. The presented data were obtained by a
semi-analytical simulation method. The results show that four
qualitatively distinguishable nutation processes influence the
oscillation of the transition rates which reduce to one
significant contribution in the cases of large and small Larmor
separations. Simple empirical equations for the calculation of
these nutation frequencies have been obtained which match the
simulated data excellently and a qualitative picture for the
interpretation of the nutation intensities has been discussed. The
presence of the four nutation processes implies that changing the
Larmor separation or the applied excitation field leads to an
abrupt and not continuous change of the observed nutation
frequencies.

\end{document}